\documentclass[twocolumn,showpacs,pra,aps,superscriptaddress]{revtex4-1}
\usepackage[utf8]{inputenc}
\usepackage{array}
\usepackage{amsmath}
\usepackage{graphicx}
\usepackage{epstopdf}
\usepackage{color}
\usepackage{amsfonts}
\usepackage{subfigure}
\usepackage{amssymb}
\usepackage{graphicx}
\usepackage{makecell}
\usepackage[colorlinks,citecolor=blue]{hyperref}

\def\ket#1{\mathinner{|{#1}\rangle}}

\newcommand{\onecolm}{
  \end{multicols}
  \vspace{-3.5ex}
  \noindent\rule{0.5\textwidth}{0.1ex}\rule{0.1ex}{2ex}\hfill
}
\newcommand{\twocolm}{
  \hfill\raisebox{-1.9ex}{\rule{0.1ex}{2ex}}\rule{0.5\textwidth}{0.1ex}
  \vspace{-4ex}
  \begin{multicols}{2}
}

\begin{document}
\title{Emergent phase transition in Cluster Ising model with dissipation}
\date{\today}

\author{Zheng-Xin Guo}
\altaffiliation{These authors contributed equally.}
\affiliation{Guangdong-Hong Kong Joint Laboratory of Quantum Matter,
Frontier Research Institute for Physics, South China Normal University, Guangzhou 510006, China}
\affiliation{Wilczek Quantum Center and Key Laboratory of Artificial Structures and Quantum Control, School of Physics and Astronomy, Shanghai Jiao Tong University, Shanghai 200240, China}

\author{Xue-Jia Yu}
\altaffiliation{These authors contributed equally.}
\affiliation{International Center for Quantum Materials, School of Physics, Peking University, Beijing 100871, China}

\author{Xi-Dan Hu}
\affiliation{Guangdong Provincial Key Laboratory of Quantum Engineering and Quantum Materials, SPTE, South China Normal University, Guangzhou 510006, China}
\affiliation{GPETR Center for Quantum Precision Measurement, South China Normal University, Guangzhou 510006, China}

\author{Zhi Li}
\email{ lizphys@m.scnu.edu.cn}
\affiliation{Guangdong-Hong Kong Joint Laboratory of Quantum Matter,
Frontier Research Institute for Physics, South China Normal University, Guangzhou 510006, China}

\begin{abstract}
We study a cluster Ising model with non-Hermitian external field which can be exactly solved in the language of free fermions. By investigating the second derivative of energy density and fidelity, the possible new critical points are tentatively located. String order parameter and staggered magnetization are then detected to reveal emergent phases of brand new characteristics. To categorize the exotic phases and phase transitions induced by non-Hermiticity, we calculate the variation mode of spin correlation function as well as string correlation function, which characterize the emergent phases and critical points with different patterns of decay and critical exponents. With the help of string order parameter and staggered magnetization, we find that there are four phases after introducing the non-Hermiticity---the cluster phase, the gapless phase, the paramagnetic (PM) phase and the antiferromagnetic (AF) phase. A phase diagram is then presented to graphically illustrate, based on two `KT-like' phase transitions and an Ising phase transition, respectively, the generation of three critical lines as non-Hermitian strength increases. Our theoretical work is expected to be realized in the experiment of ultra-cold atoms, pushing for progress in exploring novel phases and phase transitions.
\end{abstract}

\maketitle

\section{Introduction}
%%phase transition theory
According to conventional Landau-Ginzberg-Wilson paradigm~\cite{LDLandau1980}, an ordered phase is characterized by certain symmetry breaking of the system, whereas a disordered phase features the preserve of symmetry. The phase transition between ordered phases and disordered phases can be detected by certain local parameters. However, Kosterlitz and Thouless discovered a continuous phase transition without symmetry breaking in systems involving classical vortex topology, which is then named as Kosterlitz-Thouless (KT) phase transition~\cite{JKosterlitz1973}. The key characteristic of KT phase transition is that it occurs between a disordered phase and a gapless phase with quasi-long-range order which can be detected by the power-law decay of correlation function. 

%%the introduction of cluster Ising model
With the rapid development of quantum simulation and computation, the marriage of traditional condensed matter physics and cutting-edge experimental techniques gives birth to various new topics. In highly pure and controllable ultra-cold atom platforms, a triangle optical lattice can be set up with atoms loaded in a unique way~\cite{CBecker2010}, so as to create an equivalent three-spin ring-exchange interaction in spin system that can be mapped to a ``zig-zag chain”. Notably, the ground state of the system is cluster state~\cite{JKPachos2004},  a disordered state whose spin spacial rotational symmetry is protected. However, research papers also have it that except for three-spin ring-interaction, two-spin interaction relating to symmetry breaking state is also observed in such a system, causing competition between the two kinds of interaction. From the view of quantum information, researchers regard the two-spin interaction as a perturbation for it will damage the symmetry protected topological (SPT) cluster state, and they are curious about SPT's threshold of robustness~\cite{ACDoherty2009, SOSkrovseth2009, YCLi2011, XGWen2017, FPollmann2012, ZCGu2019}. From the perspective of condensed matter physics, the exotic continuous quantum phase transitions (QPT) between SPT phases and symmetry breaking phases are fascinating for their own sake~\cite{SSachdev1999, PSmacchia2011, CXDing2019}. Therefore, cluster Ising model is put forward as a good toy model to investigate the phase transitions in such quantum many-body systems. Generally, SPT cluster phase is protected by a $\mathbb{Z}_{2} \times \mathbb{Z}_{2}$ symmetry~\cite{WSon2011} and can be characterized by non-local string order~\cite{MPopp2005, LCampos2005} while the symmetry breaking antiferromagnetic (AFM) state can be identified by local staggered magnetization~\cite{CCChiang2019, VGrigorev2021}.

%%non-Hermitian and non-Hermitian cross cluster Ising
The above discussion is based on traditional quantum mechanics, which requires the Hermiticity of observables to ensure that their eigenvalues are real numbers. However, non-Hermitian physics has currently attracted extensive research interest as the non-Hermitian experimental techniques grow mature in a wide range of table-top experimental platforms~\cite{ZGong2018, KKawabata2019}, inclusive of ultra-cold atom system~\cite{JLi2019, LLi2020}, optical system~\cite{WChen2017, BPeng2014, ACerjan2019, BZhen2015}, nitrogen-vacancy center~\cite{YWu2019}, etc.
Many novel non-Hermitian phenomena have been detected, such as parity-time ($\mathcal{PT}$) symmetry~\cite{CMBender1998, CMBender2002, CMBender2007}, non-Hermitian skin effect~\cite{SYao2018, SYao20182}, new topological behaviors associated with exceptional points~\cite{DWZhang2020, PHe2020, HShen2018, EJBergholtz2021, AGhatak2019, KKawabata2019, YHe20202}, disorder induced by non-Hermiticity~\cite{DWZhang20202, XWLuo2020, HJiang2019}, etc. Recently, much attention has been paid on how non-Hermiticity influences quantum phase transitions in systems with rotation-time-reversal ($\mathcal{RT}$) symmetry~\cite{TELee2014, XZZhang2013, CWang2020, CLi2014, KLZhang2020}, spin model with imaginary external field~\cite{SB2021, YNishiyama2020, YNishiyama20202, YGLiu2021, NShibata2019} and so on. It is also worth mentioning that some researchers have developed a set of non-Hermitian linear response theory, which uses dissipation as a means to reveal the properties of Hermitian equilibrium system~\cite{LPan2020}.

%%summary of our work
In this work, we try to uncover the connection between novel emergent phases and non-Hermiticity without $\mathcal{PT}$ or $\mathcal{RT}$ symmetry. The main body of this paper is organized as follows. In Sec.~\ref{sec:model}, we introduce cluster Ising model with on-site dissipations and analytically transform it into free fermion expression. In Sec.~\ref{sec:observables}, the observables and the methods we adopted to characterize quantum phase transitions are presented. In Sec.~\ref{sec:result}, we give the theoretical results of observables and discuss the exotic phase diagram and phase transitions. We summarize our work in Sec.~\ref{sec:summary}.

\section{Model and Exact solution}\label{sec:model}
\begin{figure}[tbhp] \centering
\includegraphics[width=0.45\textwidth]{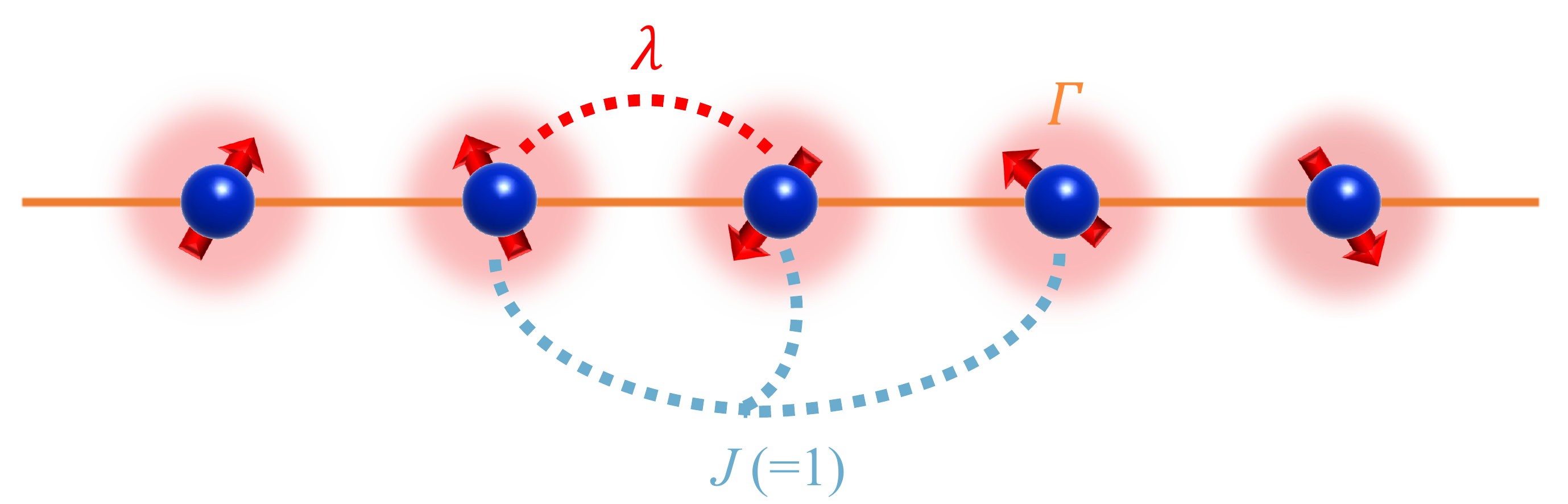}
\caption{(Color online). Graphic demonstration of cluster Ising model with dissipation, where $\lambda/J$ is the ratio of Ising exchange strength to cluster exchange strength and $\it\Gamma$ is the strength of loss or gain. We take $J$=1 in the following discussion.}\label{modeldemo}
\end{figure}

In this section, we establish the non-Hermitian cluster Ising model and conduct the diagonalization procedure to obtain exact solution of the ground state. Based on the conventional  cluster Ising model, we build up our Hamiltonian by inserting dissipation, which is equivalent to an external imaginary field. The expression reads
\begin{equation}
    H=-J\sum_{l=1}^{N}\sigma_{l-1}^{x}\sigma_{l}^{z}\sigma_{l+1}^{x}+\lambda\sum_{l=1}^{N}\sigma_{l}^{y}\sigma_{l+1}^{y}+\frac{i\Gamma}{2}\sum_{l=1}^{N}\sigma_{l}^{
    u},
\end{equation}
where $\sigma^{x}_{l}, \sigma^{y}_{l}$ and  $\sigma^{z}_{l}$ are pauli matrices of the $l^{th}$ spin and $\sigma^{u}$ denotes the matrix $\begin{bmatrix}
1 & 0  \\
0 & 0
\end{bmatrix}$ corresponding to the loss or gain, which is a simple way to involve non-Hermiticity in optical or atomic experiments~\cite{BPeng2014, XWLuo2020, Muller2012, JMZeuner2015}. $N$ is large enough for us to view $N/2$ as a `decent half' regardless of the parity of $N$. There are three control parameters $\it J$, $\lambda$ and $\Gamma$ in our model, where the former two indicate the competition between SPT phase and symmetry breaking phase while the latter determines the strength of complex field (see Fig.~\ref{modeldemo}). Notably, in this work, we set $J=1$ and take it as energy unit in the following calculations.

The diagonalization process can be divided into three steps. First of all, we rewrite the Hamiltonian in free fermionic language  using the standard Jordan-Wigner transformation, which is defined as
\begin{equation}
    \sigma_{l}^{z}=1-2c_{l}^{\dagger}c_{l},
\end{equation}
\begin{equation}
    \sigma_{l}^{+}=\prod_{j<l}(1-2c_{j}^{\dagger}c_{j})c_{l},
\end{equation}
where $c_l^{\dagger}$ and $c_l$ are the creation and annihilation operators at site $l$, respectively. Then we can obtain the Hamiltonian in spinless fermion expression as
\begin{equation}
\begin{aligned}
H=& \sum_{l}^{N}\left[-\left(c_{l}^{\dagger}-c_{l}\right)\left(c_{l+2}^{\dagger}+c_{l+2}\right)\right.\\
&\left.+\lambda\left(c_{l}^{\dagger}+c_{l}\right)\left(c_{l+1}^{\dagger}-c_{l+1}\right)\right]\\
&-\frac{i\Gamma}{2 } \sum_{l}^{N}\left(1- c_{l}^{\dagger} c_{l}\right).
\end{aligned}
\end{equation}
Secondly, a Fourier transformation $c_{l}=\frac{1}{\sqrt{N}}\sum_{k=-\pi/2}^{\pi/2}$\\$e^{2\pi ikl/N} b_k$ is conducted and we have
\begin{equation}
\begin{aligned}
 H=&2 \sum_{k}\left[i y_{k}\left(c_{k}^{\dagger} c_{-k}^{\dagger}+c_{k} c_{-k}\right)\right.\\
&\left. +z_{k}\left(c_{k}^{\dagger} c_{k}+c_{-k}^{\dagger} c_{-k}-1\right)\right].
\end{aligned}
\end{equation}
Here, $y_k=-\rm{sin}(2k)-\lambda \rm{sin}(k)$ and $z_k=-\rm{cos}(2\it k)+\lambda \rm{cos}(\it k)+\frac{i\Gamma}{4}$. Thirdly, a Bogoliubov transformation helps diagonalize the above equation, which reads
\begin{equation}
    b_{k}=\cos(\frac{\theta_k}{2})\gamma_{k}+i\sin(\frac{\theta_k}{2})\gamma_{-k}^{\dagger},
\end{equation}

\begin{equation}
    b_{-k}=\cos(\frac{\theta_k}{2})\gamma_{-k}-i\sin(\frac{\theta_k}{2})\gamma_{k}^{\dagger}.
\end{equation}
Eventually, a diagonalized solution is acquired as
\begin{equation}
    H=\sum_{k>0} \Lambda_{k}(\gamma_k^{\dagger}\gamma_{k}-\frac{1}{2}),
\end{equation}
where
\begin{equation}
    \Lambda_k=\sqrt{y_{k}^2+z_{k}^2},
\end{equation}
\begin{equation}
{\rm{tan}}(\theta_k)=-\frac{y_k}{z_k},
\end{equation}
and the ground state of the model is
\begin{equation}
    \ket{G}=\prod_{k>0}[{\rm{cos}}(\frac{\theta_k}{2})+i{\rm{sin}}(\frac{\theta_k}{2})c^{\dagger}_k c^{\dagger}_{-k}]\ket{{\rm{Vac}}},
\end{equation}
where $\ket{{\rm{Vac}}}$ denotes the vacuum state of the free fermion.

\section{Observables and methods}\label{sec:observables}
\subsection{Ground state energy density and its second derivative}

The non-analyticity in the second derivative of ground state energy density implies that continuous QPT occurs at zero temperature. According to the above solution, the ground state energy can be calculated analytically and numerically via the equation
\begin{equation}
U_g=-\frac{2}{N} \sum_{k>0} \sqrt{y_{k}^{2}+z_{k}^{2}}=-\frac{1}{\pi} \int_{0}^{\pi} \sqrt{y_{k}^{2}+z_{k}^{2}} d k,
\end{equation}
and we can easily obtain the second derivative of $U_g$ with respect to $\lambda$, i.e., $\frac{\partial^2 U_g}{\partial \lambda^2}$.

\begin{figure}[tbhp] \centering
\includegraphics[width=0.45\textwidth]{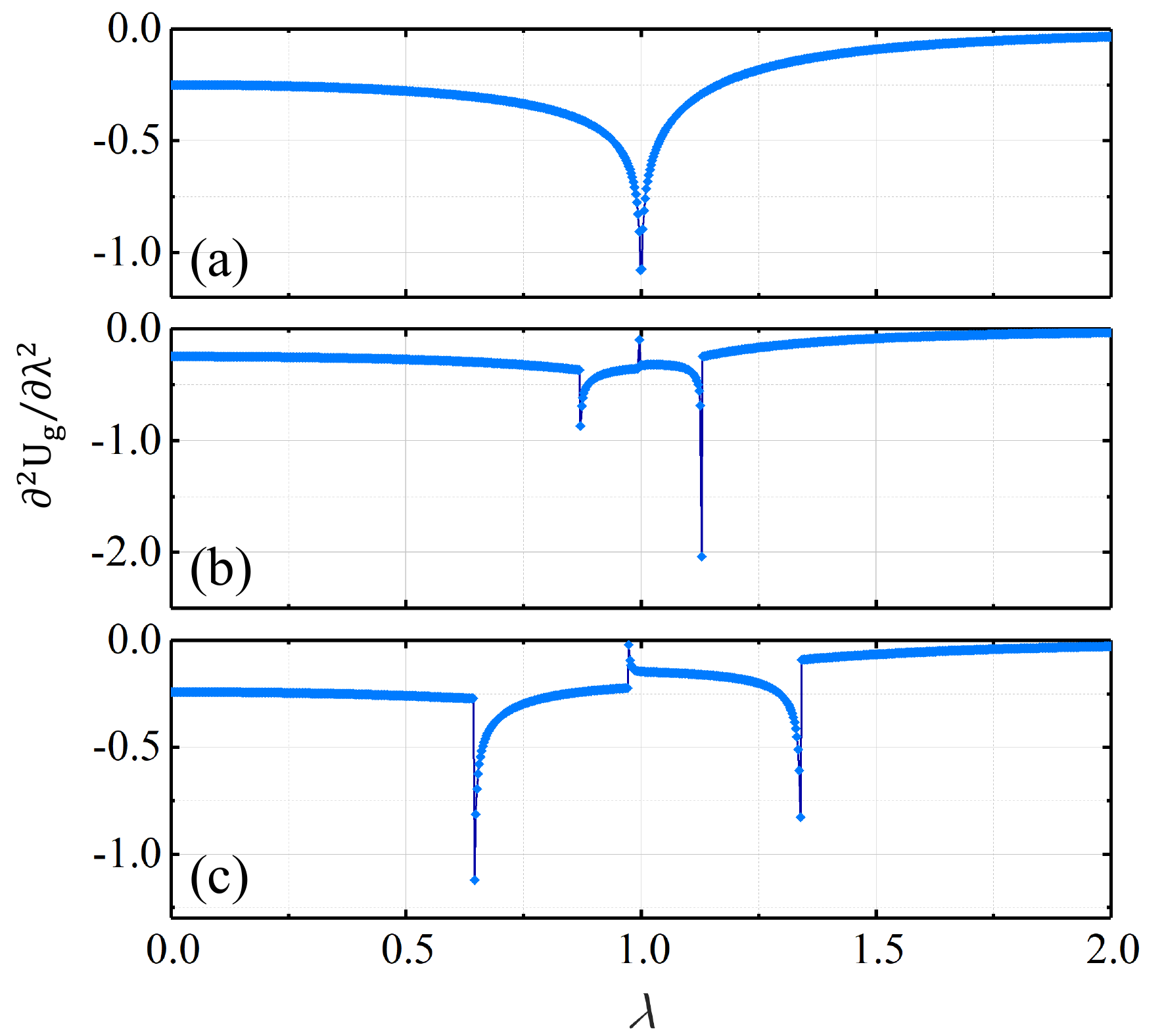}
\caption{(Color online). The second derivative of ground state energy density with respect to $\it\lambda$. (a) When $\it\Gamma=$0, the singularity occurs at $\it\lambda=$1, which is in good agreement with the critical point of standard clusing-Ising model. (b) When $\it\Gamma=$0.6, the non-analytical point in Hermitian case splits into 3 points, indicating new potential critical points. (c) When $\it\Gamma=$1.6, distance between the three non-analytical points increases with the growing non-Hermitian strength.}\label{ugderivative}
\end{figure}

\subsection{Fidelity}
As the inner product of two wave functions with a tiny difference in parameters, fidelity is also efficient in indicating the critical points of QPT, whose expression reads
\begin{equation}
F(\lambda,\lambda+\epsilon)=\langle\mathrm{G}(\lambda) \mid \mathrm{G}(\lambda+\epsilon)\rangle=\prod_{k>0} F_{k},
\end{equation}
with
\begin{equation}
F_{k}=\cos \frac{\theta_{k}(\lambda)-\theta_{k}(\lambda+\epsilon)}{2}.
\end{equation}
It is noticeable that  $\epsilon$ should take a small value. However, some arbitrariness in choosing the value of $\epsilon$ can be allowed within a reasonable range, and the selection rule is just as what we have discussed before~\cite{CXDing2019}. Here, we set  $\epsilon=10^{-5}$, which can ensure the stability of results under different parameters.

\subsection{Order parameters}
Two kinds of order parameters will be investigated in this section, i.e., non-local string order parameter characterizing disordered phase, and local staggered magnetization characterizing AFM phase. These order parameters will be non-zero as long as the system is in the corresponding phase.

\begin{figure}[tbhp] \centering
\includegraphics[width=0.45\textwidth]{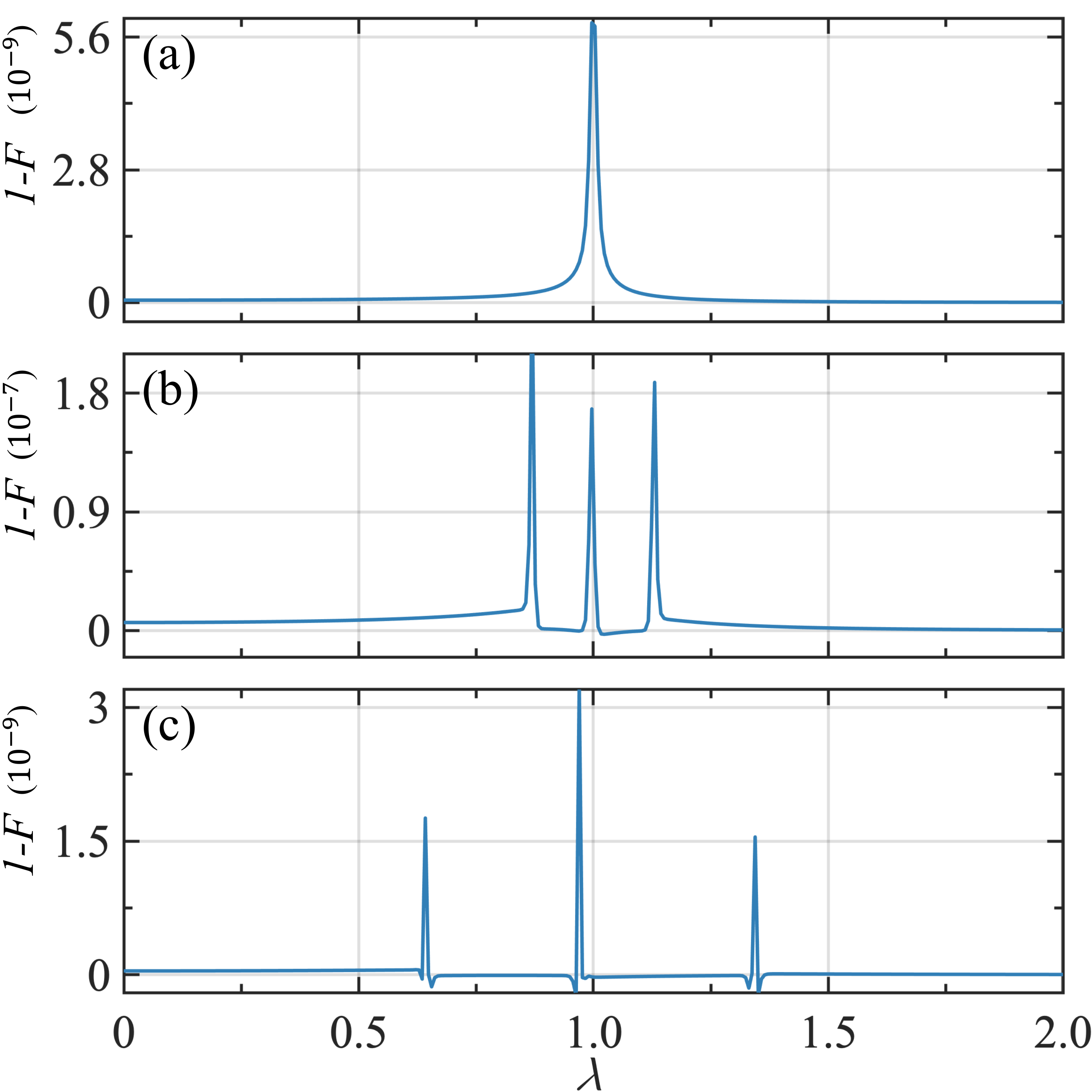}
\caption{(Color online). The distribution of $1-F$ (fidelity) versus $\lambda$. (a) When $\it\Gamma=$0, the singular point well indicates the critical point  of conventional SPT-AFM phase transition. (b) When $\it\Gamma=$0.6, the fidelity also characterizes possible new phase transition points. (c) As $\it\Gamma$ ascends to 1.6, the behavior of singularity is the same as that of  the second derivative of energy density.}\label{fidelity}
\end{figure}

Let us begin with the string order parameter, which is defined as
\begin{equation}
\mathcal{O}^{x}=\lim _{N \rightarrow \infty}(-1)^{N}\left\langle\sigma_{1}^{x} \sigma_{2}^{y}\left(\prod_{k=3}^{N-2} \sigma_{k}^{z}\right) \sigma_{N-1}^{y} \sigma_{N}^{x}\right\rangle_{0}.
\end{equation}
Using the technique in Ref.~\cite{GCWick1950}, we can express it by the product of $A_j=c_j^{\dagger}+c_j$ and $B_j=c_j-c_j^{\dagger}$, i.e.,
\begin{equation}
\mathcal{O}^{x}=\lim _{r \rightarrow \infty}\left\langle B_{2} A_{3} B_{3} \ldots A_{r} B_{r} A_{r+1} B_{r+1} A_{r+2}\right\rangle.
\end{equation}
Then, with the help of Wick theorem~\cite{GCWick1950},  we can go on expanding it by the contractions $\langle A_jA_l\rangle$, $\langle B_jB_l\rangle$ and $\langle B_jA_l\rangle$, whose expression can be acquired using the ground state function. We have
\begin{equation}
    \langle A_j A_l \rangle=\delta_{jl},
\end{equation}

\begin{equation}
    \langle B_j B_l \rangle=-\delta_{jl},
\end{equation}

\begin{equation}
\begin{aligned}
\langle B_{j} A_{l}\rangle&=G_{j, l}=G_{r}\\
&= \frac{1}{\pi} \int_{0}^{\pi} d k[\cos (k r) \cos \theta_{k} +\sin (k r) \sin \theta_{k}],
\end{aligned}
\end{equation}
where $r=j-l$. Since $\langle A_jA_l\rangle$ and $\langle B_jB_l\rangle$ are always equal to zero and by considering the characteristics of their expression, $\mathcal{O}^{x}$ can be transformed to a Toeplitz determinant,

\begin{equation}\label{stringcorrel}
\mathcal{O}^{x}=\lim _{r \rightarrow \infty}\left|\begin{array}{cccc}
G_{-2} & G_{-3} & \cdots & G_{-r-1} \\
G_{-1} & G_{-2} & \cdots & G_{-r} \\
G_{0} & G_{-1} & \cdots & G_{-r+1} \\
\vdots & \vdots & \ddots & \vdots \\
G_{r-2} & G_{r-5} & \cdots & G_{-3}\\
G_{r-3} & G_{r-4} & \cdots & G_{-2}
\end{array}\right| .
\end{equation}
From the perspective of numerical calculation, we can take finite $r$ and calculate finite number of integrals to obtain $\mathcal{O}^{x}$. Note that, $r$ should be as large as possible in order to approach the thermodynamic limit.

Next, we can acquire staggered magnetization at temperature $T$ via the calculation of spin correlation function
\begin{equation}
R_{j l}^{\alpha}(T)=\left\langle\sigma_{j}^{\alpha} \sigma_{l}^{\alpha}\right\rangle_{T},
\end{equation}
where $\alpha$ can be $x$, $y$ or $z$. Let us take $R_{j l}^{x}(T)$ as an example. Similar to the precedure of calculating string correlation function, $R_{j l}^{x}(T)$ can also be expanded by $ A_{j}=c_{j}^{\dagger}+c_{j}, B_{j}=c_j-c_j^{\dagger}$, that is,
\begin{equation}
\begin{aligned}
R_{j l}^{x}(T) &=\left\langle\left(c_{j}-c_{j}^{\dagger}\right) \prod_{j<m<l}\left(1-2 c_{m} c_{m}^{\dagger}\right)\left(c_{l}^{\dagger}+c_{l}\right)\right\rangle_{T} \\
&=\left\langle B_{j} A_{j+1} B_{j+1} \ldots A_{l-1} B_{l-1} A_{l}\right\rangle_{T}.
\end{aligned}
\end{equation}
\begin{figure}[tbhp] \centering
\includegraphics[width=0.45\textwidth]{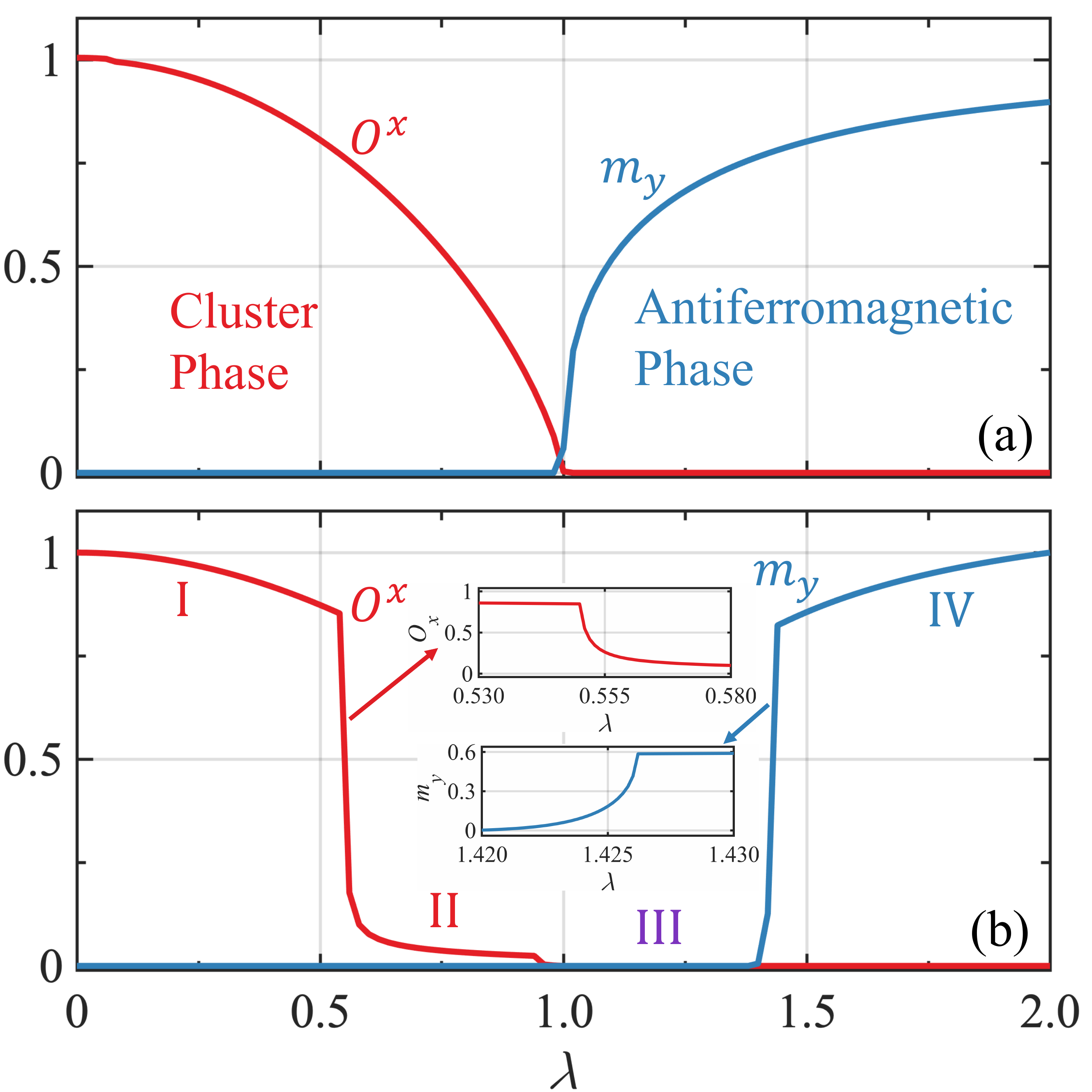}
\caption{(Color online). The numerical results of string correlation function $\it O_x$ and staggered magnetization $\it m_y$ for (a) Hermitian case with $\it \Gamma$=0 and (b) non-Hermition case with $\it \Gamma$=2.0. The insets of (b) are close-ups of the continuous change region around the two phase transition points.}\label{orderparam}
\end{figure}By using Wick theorem again we can also convert $R_{j l}^{x}(T)$ into a Toeplitz determinant
\begin{equation}
R_{r}^{x}(T)=\left|\begin{array}{cccc}
D(-1, T) & D(-2, T) & \cdots & D(-r, T) \\
D(0, T) & D(-1, T) & \cdots & D(-r+1, T) \\
\vdots & \vdots & \ddots & \vdots \\
D(r-2, T) & D(r-3, T) & \cdots & D(-1, T)
\end{array}\right|,
\end{equation}
where the elements of the determinant reads
\begin{equation}
\left\langle B_{j} A_{l}\right\rangle_{T}=D_{j l}(T)=D(j-l, T)=D(r, T).
\end{equation}
Similarly, $R_{r}^{y}(T)$ can be expressed by

\begin{equation}
R_{r}^{y}(T)=\left|\begin{array}{cccc}
D(1, T) & D(0, T) & \cdots & D(-r+2, T) \\
D(2, T) & D(1, T) & \cdots & D(-r+3, T) \\
\vdots & \vdots & \ddots & \vdots \\
D(r, T) & D(r-1, T) & \cdots & D(1, T)
\end{array}\right|.
\end{equation}
With $R_{r}^{y}$, we can calculate staggered magnetization $m_y$ with the definition
\begin{equation}\label{staggeredma}
\lim _{r \rightarrow \infty}(-1)^{r} R_{r}^{y}(0)=m_{y}^{2}.
\end{equation}

Overall, with the above expression, we can investigate the influence of non-Hermiticity on the CIM's phase distribution at certain temperature.

\begin{figure*}[tbhp] \centering
\includegraphics[width=1\textwidth]{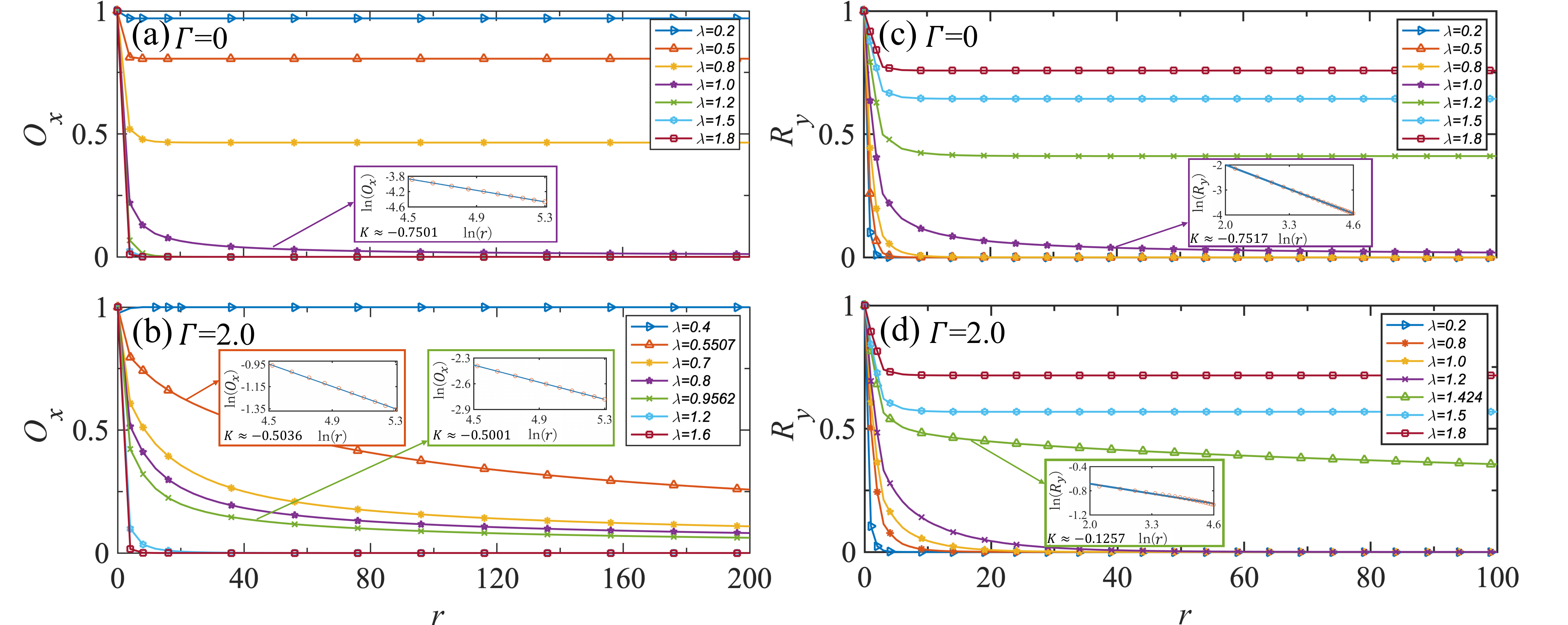}
\caption{(Color online). 
The variation modes of string parameter $|O_x(r)|$ and correlation function $|R_y(r)|$ at different $\lambda$ for (a) and (c) Hermitian case with $\it\Gamma$=0 as well as (b) and (d) non-Hermitian case with $\it\Gamma$=2.
When $\it\Gamma$=0, the insets plot the curves at critical point $\lambda=$1 in ln-ln coordinates and show the slope $K$ of the lines. As $\it\Gamma$ increases to 2, the three critical points are at 0.551, 0.956 and 1.424. The insets also show the curves featuring power-law decay in ln-ln coordinates (with subscript $e$).}\label{spincf}
\end{figure*}

\subsection{Variation mode of correlation function}

In phase transition theory, the spin correlation function $R_y(r)$ decreases to a fixed non-zero value as $r$ increases in symmetry breaking phase, in contrast with the exponential decay to zero in disordered phase. At the critical point between these two phases, $R_y(r)$ exhibits a unique power-law decay. Similarly, the variation mode of string order parameter $O_x(r)$ also carry the key information of the phases and phase transitions. 
$O_x(r)$ tends to be a constant in non-trivial cluster phase, rather than the exponential decay in trivial phases.
 Moreover, it is worth mentioning that if one plot the curve of power-law decay in the ln-ln coordinates, it will become a straight line and the slope of the line will directly reveal one of the critical exponents $\eta$.

\section{Results and Discussions}\label{sec:result}

In this section, we are going to illustrate the influence of non-Hermiticity on different observables. Let us start with the second derivative of ground state energy density. In Hermitian case, as it is shown in Fig.~\ref{ugderivative} (a), the singularity emerges at $\lambda=1$, which corresponds to the critical point of standard cluster Ising model. However, when non-Hermitian strength increases [see Fig.~\ref{ugderivative} (b) and (c)], the singular point turns into three points, which drift apart from one another with an increasing $\it \Gamma$. The emergence of new non-analytical points indicates possible new phase transitions that remains unknown. Notably, the continuous first derivative of $U_g$ is also shown in Appendix A to prove that all these three phase transitions are continuous phase transitions.

We also demonstrate the fidelity whose singularity characterizes phase transition as well. As is shown in Fig.~\ref{fidelity} (a), in Hermitian case, the singular point also works well in characterizing the SPT-AFM phase transition at $\lambda=1$. Then, we turn to the non-Hermitian cases as $\it\Gamma$ increases to $0.6$ and $1.6$ [see Fig.~\ref{fidelity} (b) and (c)]. The behavior of singularity is in good agreement with that of $\partial^2 U_g/\partial \lambda^2$. We can observe two more emerging singular points and witness them moving away from each other as non-Hermiticity increases, which denotes the emergence of the unknown phase transition. The results of fidelity and the second derivative of ground state energy nicely support each other.

When one investigate Hermitian cluster Ising model, it is a standard procedure to calculate the string order parameter $\it O_x$ and staggered magnetization $\it m_y$. One can directly recognize the domination of disordered (ordered) phase with the help of non-local (local) order parameter string order parameter (staggered magnetization) provided it is non-zero. We also investigate the distribution of these two order parameters under different non-Hermitian strengths. In Hermitian case [see Fig.~\ref{orderparam} (a)], the behaviors of $\it O_x$ and $\it m_y$ are the same as the previous researches, where $\it O_x$ ($\it m_y$) is non-zero when $\lambda<1$ ($\lambda>1$) and its decline is continuous. However, when we increase the non-Hermitian strength $\it\Gamma$ from 0 to 2, the distribution of order parameters changes greatly. At first, on the one hand, $\it O_x$ shows an intense but continuous decline after the kink at the first critical point, while on the other hand, the parameter $\it m_y$ illustrates a sharp but continuous increase before the kink at the third critical point. The insets of Fig.~\ref{orderparam} (b) show the close-ups of continuous variation around the two kinks at critical points. Note that, although the kinks at continuous critical points are not to be found in Hermitian case, recent studies have proved that this kind of kinks is actually a real phenomenon in the non-Hermition case.~\cite{TELee2014, BBWei2017}. Thus, after involving dissipation strength, one can witness that the original two phase zones are divided into four. Then, one can see that in zone \uppercase\expandafter{\romannumeral1} and \uppercase\expandafter{\romannumeral2}, $\it O_x$ is non-zero and $\it m_y$ is zero. In zone \uppercase\expandafter{\romannumeral4}, $\it m_y$ is non-zero and $\it O_x$ is zero. However, in zone \uppercase\expandafter{\romannumeral3}, both $\it O_x$ and $\it m_y$ are zero, which is beyond the way of phase classification in Hermitian case. Besides, it is noticeable that this is a numerical result where $\it r$ in Eq.~(\ref{stringcorrel}) and Eq.~(\ref{staggeredma}) is taken as $1000$ instead of infinite. Thus, tiny numerical error is involved, leading to some continuous transition areas between different zones. In principle, there will be some non-analytical slump as long as $\it r$ is large enough to approach the thermodynamics limit, the position of which is also consistent with that of singular points as we calculated above.

From the above discussions, it is obvious that the introduction of non-Hermitian term leads to new phases and phase transitions that transcend the framework of Hermitian case. After that, we want to classify the emergent phases and phase transitions with the help of the variation modes of spin correlation function $R_y(r)$ and string parameter $O_x(r)$. 
We investigate the variation of $O_x( r)$ and $R_y(r)$ at different $\lambda$ and under different non-Hermitian strengths $\it\Gamma$. When $\it\Gamma$ = 0, the critical point is at $\lambda$ = 1, which denotes the cluster-AFM phase transition with symmetry breaking. From Fig.~\ref{spincf} (a), we can see that $O_x( r)$ remains a constant value before $\lambda =$ 1 and shows exponential decay to zero after $\lambda =$ 1, indicating non-trivial cluster phase and trivial AFM phase, respectively. Meanwhile, from Fig.~\ref{spincf} (c), we can see that the curves of $R_y(r)$  before $\lambda =$ 1 exhibit the exponential decay and those after $\lambda=$ 1 decrease to fixed non-zero values, denoting symmetry protected cluster phase and symmetry breaking AFM phase, respectively. Notably, as is shown in the insets of the two graphs, the curves at the critical point show the power-law decay, indicating the phase transition. 
The critical exponent $\eta$ at that critical point can also be obtained as 3/4 according to the slope of line in the ln-ln coordinates (natural logarithm), which is different from that of Ising universality class. 
Previous work~\cite{PSmacchia2011} also points out that although most of the critical exponents of Hermitian cluster Ising model equal to that of Ising universality class, the central charge of them are different, which means phase transition in Hermitian cluster Ising model is beyond Ising universality class. From our calculation, the difference in critical exponent $\eta$ also provides more evidence to support the above claim. 

Interestingly, as $\it \Gamma$ increases to 2 [see Fig.~\ref{spincf} (b) and (d)], exotic phases and phase transitions are revealed by the variation mode of $O_x(r)$ and $R_y(r)$. Let us start with $O_x( r)$ [see Fig.~\ref{spincf} (b)]. In zone \uppercase\expandafter{\romannumeral1} before the first critical point,  $O_x( r)$ remains a constant as string length $\it r$ increases, which indicates it is a non-trivial cluster phase. In zone \uppercase\expandafter{\romannumeral3} and \uppercase\expandafter{\romannumeral4}, $O_x(r)$ decreases exponentially to zero, which indicates they are trivial phases. However, in the whole zone \uppercase\expandafter{\romannumeral2}, $O_x(r)$ 
exhibits a power-law decay and the slope of curves in ln-ln coordinate system is close to -0.5, which indicates that zone \uppercase\expandafter{\romannumeral2} is a gapless phase and the critical exponent $\eta$ of the phase transitions is 1/2. 
Then, let us move on to the behavior of $R_y(r)$ [see Fig.~\ref{spincf} (d)]. One can only observe the power-law decay at the third critical point, which means that the symmetry breaking only occurs at the third phase transition. Analyzing the above two clues with the distributions of $\it O_x$ and $\it m_y$ shown in Fig.~\ref{orderparam} (b), we find that in zone \uppercase\expandafter{\romannumeral4}, $\it O_x$ is zero and AFM order parameter $\it m_y$ is non-zero, which means that zone \uppercase\expandafter{\romannumeral4} is a trivial AFM phase with symmetry breaking. However, both $\it O_x$ and $\it m_y$  are zero in zone \uppercase\expandafter{\romannumeral3}, which means it is a trivial phase without symmetry breaking and magnetism. Thus, it can only be a PM phase in one dimensional chain. Therefore, it should be a PM-AFM phase transition of standard Ising universality class between zone \uppercase\expandafter{\romannumeral3} and \uppercase\expandafter{\romannumeral4}. This claim is also supported by the critical exponent obtained by the slope of $R_y(r)$ at the critical point between zone \uppercase\expandafter{\romannumeral3} and \uppercase\expandafter{\romannumeral4}: $\eta=$1/8, which is consistent with that of standard Ising universality class. 
As a result, the phase transition between zone \uppercase\expandafter{\romannumeral1} and \uppercase\expandafter{\romannumeral2} as well as zone \uppercase\expandafter{\romannumeral2} and \uppercase\expandafter{\romannumeral3} are phase transitions between gapped phase and gapless phase, so we call them `KT-like' phase transition.
To conclude, the introducing of non-Hermitian term gives rise to novel phases and phase transitions as well as the shift of critical points.

\begin{figure}[tbhp] \centering
\includegraphics[width=0.45\textwidth]{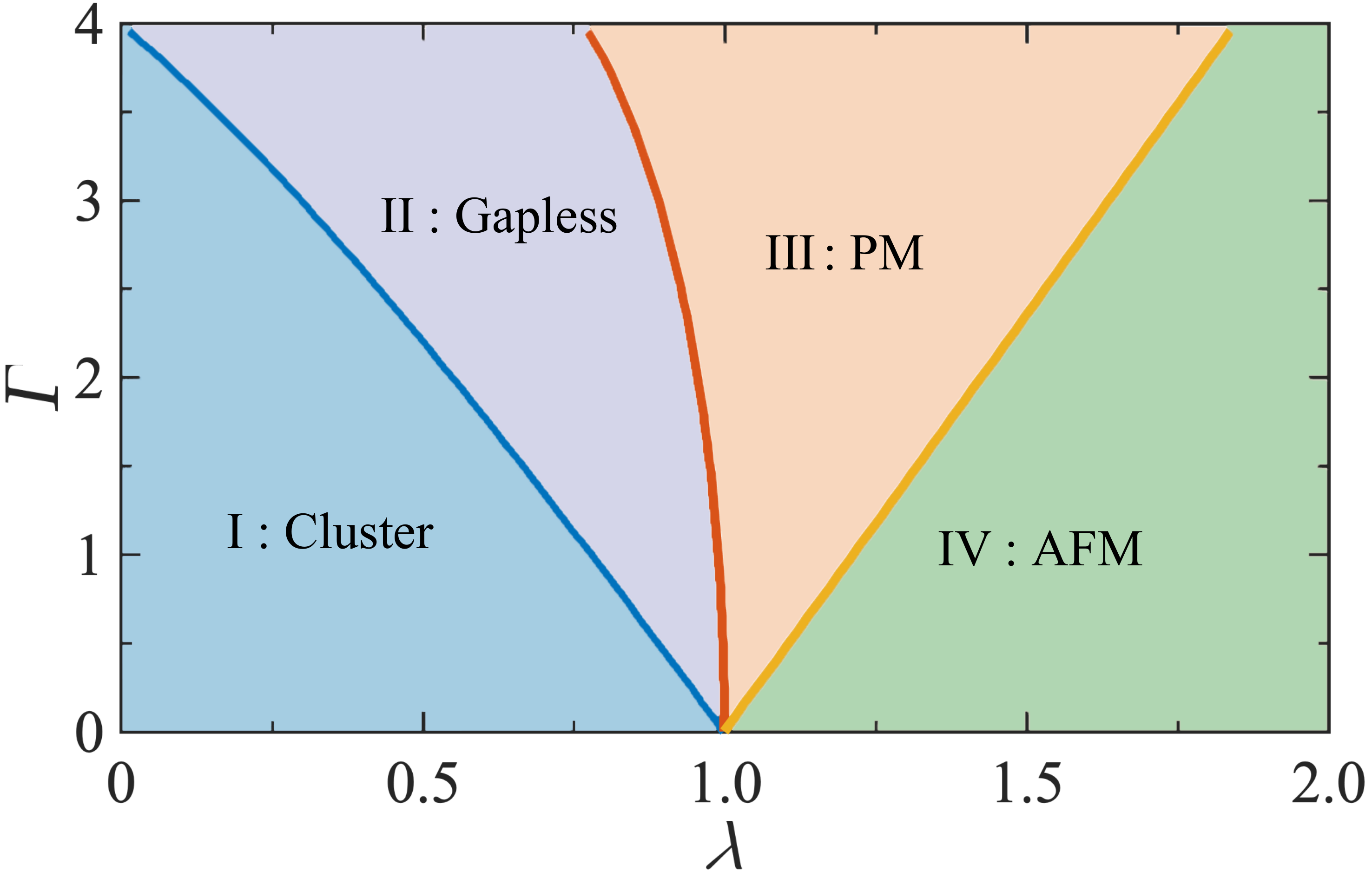}
\caption{(Color online). The $\it\Gamma-\lambda$ phase diagram according to $\it U_g$'s second derivative. In Hermitian case ($\it\Gamma=$ 0), there is a conventional critical point of Ising universal class at $\lambda=$1. With the increase of $\it\Gamma$, the critical point splits into three critical points, denoting the phase transitions among the four different phases.}\label{phasediag}
\end{figure}

To be more comprehensive and intuitive, we investigate the $\it\Gamma-\lambda$ phase diagram according to the non-anlytical points of $\it U_g$'s second derivative (see Fig.~\ref{phasediag}). When $\it\Gamma=$ 0, there is only one SPT-AFM phase transition that is beyond Ising universality class at $\lambda=$1. As $\it\Gamma$ grows, the critical point extends to three consecutive critical lines. Blue, red and yellow lines correspond to two `KT-like' phase transition and a standard Ising phase transition, respectively. Notably, only phase transition on yellow critical line still accompanies with the breaking of spin rotational symmetry.

\section{Summary}\label{sec:summary}
There is an SPT phase and an AFM phase in Hermitian cluster Ising model and the phase transition between them is beyond Ising universality class. The influence of non-Hermiticity on cluster Ising model is investigated in this work. We first detect the singular behaviors of second derivative of energy density and fidelity, finding that new critical points may emerge and move away from each other. Next, we numerically investigate the string order parameter and staggered magnetization, and the results show that the introduction of non-Hermiticity will give rise to a very rich phase diagram featuring phase transitions between four phases. In order to characterize them, we then investigate the variation modes of string parameter and spin correlation function, which help us distinguish different phases and characterize three phase transitions with the critical exponent. By combining the result of string order parameter and staggered magnetization, the four phases are identified as a cluster phase, a gapless phase, a paramagnetic phase and an antiferromagnetic phase, respectively. The phase transitions between them are two `KT-like' phase transitions and one standard Ising phase transition with the corresponding critical exponents $\eta=$1/2, 1/2 and 1/8, respectively. Eventually, we give the $\it\Gamma-\lambda$ phase diagram to visualize the emergence and extension of critical lines.

In the end, we briefly clarify the possible mechanism behind the novel phenomena. In Hermitian case, when $\lambda<$1, SPT phase of the cluster Ising model is well protected by a $\mathbb{Z}_{2} \times \mathbb{Z}_{2}$ symmetry, i.e., both the $Z2$ symmetry of $\sigma_z$'s product and the symmetry of time-reversal are unbroken. However, after introducing the non-Hermitian external field, though the $\mathbb{Z}_{2}$ symmetry of $\sigma_z$'s product is intact, the time-reversal symmetry is broken, which leads to the emergence of cluster phase in zone \uppercase\expandafter{\romannumeral1}, the closing of gap in zone \uppercase\expandafter{\romannumeral2} as well as the emergence of paramagnetic phase in zone \uppercase\expandafter{\romannumeral3}. Deeper mechanism behind the relation between non-Hermitian external field and gapless phase remains unclear nowadays, which is related to the developing non-unitary conformal field theory and we will discuss it in our future work. Our work can be realized in ultra-cold atom experiment and will shed light on experimental construction of novel phases and phase transitions in open quantum many body systems.\\

\section*{Acknowledgment}
We thank C. X. Ding, Y. G. Liu, Q. Q. Su, M.Gong, S. Liu and D. C. Lu for helpful discussions. This work was supported by the NSFC (Grant No. 11704132), the NSAF (Grant No. U1830111), the Natural Science Foundation of Guangdong Province (Grant No. 2021A1515012350), and the KPST of Guangzhou (Grant No. 201804020055).
\\

\appendix
\section{First derivative of ground state energy density}

\begin{figure}[tbhp] \centering
\includegraphics[width=0.48\textwidth]{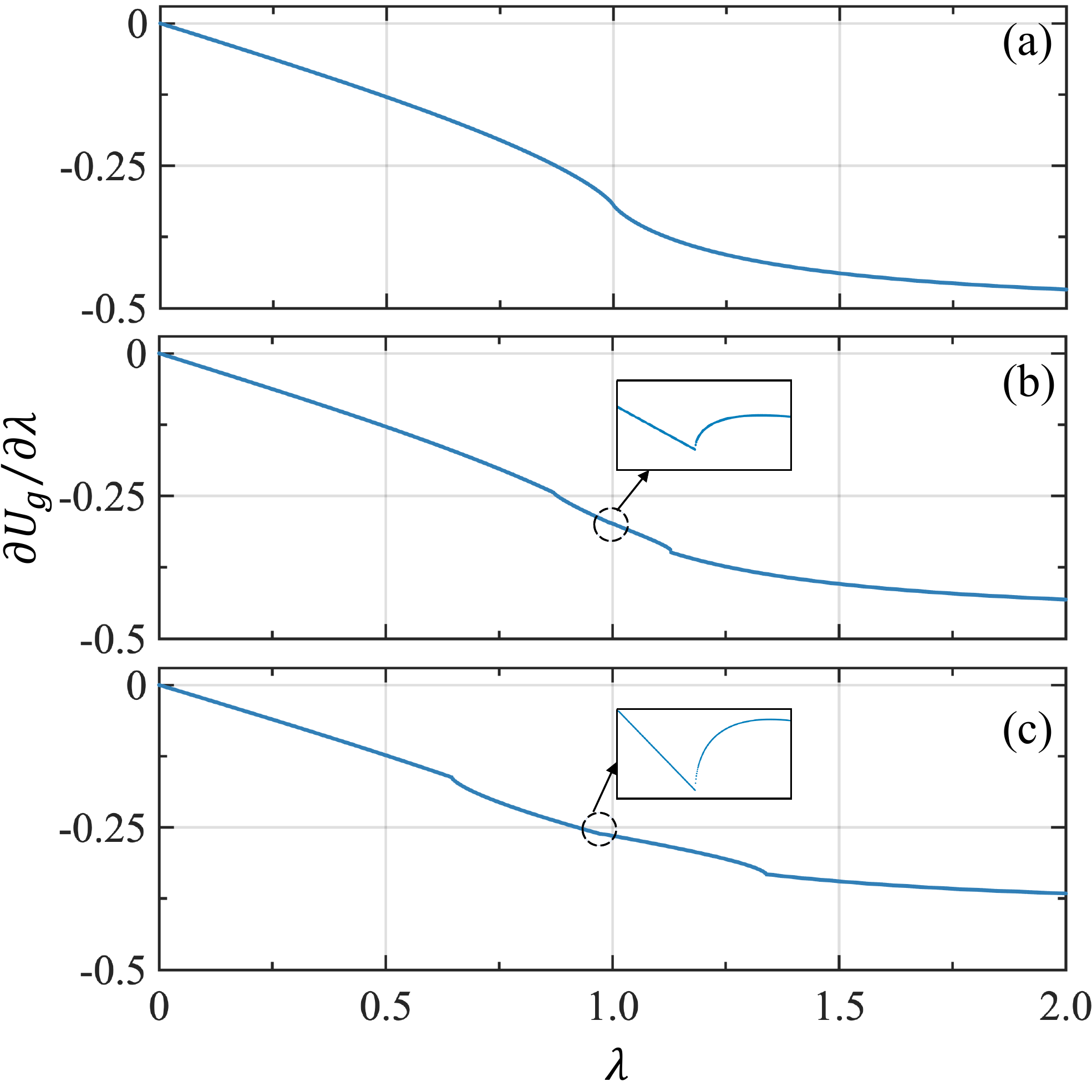}
\caption{(Color online). The first derivative of ground state energy density $U_g$ under different dissipation strength $\it \Gamma$. (a) When $\it \Gamma=$ 0, the distribution of $\partial U_g/\partial \lambda$ with $\lambda$. (b) When $\it \Gamma=$ 0.6, the distribution of $\partial U_g/\partial \lambda$ with $\lambda$. (c) When $\it \Gamma=$ 1.6, the distribution of $\partial U_g/\partial \lambda$ with $\lambda$. The insets: the zoom in of the central area where the second critical point at. All of the above distributions are continuous at the critical points.}\label{FirstUg}
\end{figure}

To prove that all the phase transitions in non-Hermitian case are continuous, we present the first derivative of the ground state energy density $U_g$ (see Fig.~\ref{FirstUg}), from which one can see that the first derivative of the order parameters is continuous. Since the second derivative of $U_g$ are discontinuous (see Fig.~\ref{ugderivative}), all the phase transitions after introducing non-Hermition dissipation are standard continuous phase transitions.

%%bibliography

\end{document}